\documentclass[manuscript]{aastex}

\slugcomment{To appear in Astrophysical Journal}
\shorttitle{Fermi J1049.7 and J1103.2+1145}
\shortauthors{Nishimichi et al.}

\begin{document}

\title{Fermi LAT detection of two high Galactic latitude gamma-ray sources, Fermi J1049.7+0435 and J1103.2+1145}
\author{Masaki Nishimichi, Takeshi Okuda and Masaki Mori}
\affil{Department of Physical Sciences, Ritsumeikan University, Kusatsu, Shiga 525-8577, Japan}
\author{Philip G. Edwards}
\affil{CSIRO Astronomy and Space Science, PO Box 76, Epping NSW 1710, Australia}
\author{Jamie Stevens}
\affil{CSIRO Astronomy and Space Science, Locked Bag 194, Narrabri NSW 2390, Australia}

\begin{abstract}

During a search for gamma-ray emission from NGC 3628 (Arp 317), 
two new unidentified gamma-ray sources, Fermi J1049.7+0435 and J1103.2+1145 have
been discovered \citep{ATel}.
The detections are made in data from the Large Area Telescope (LAT), 
on board the Fermi Gamma-ray Space Telescope, 
in the 100\,MeV to 300\,GeV band during the period between 2008 August 5 and 2012 October 27. 
Neither is coincident with any source listed in the 2FGL catalogue \citep{Nolan2012}. 
Fermi J1049.7+0435 is at Galactic coordinates 
$(l,b) = (245.34^\circ , 53.27^\circ )$, 
$(\alpha_{J2000} , \delta_{J2000}) = (162.43^\circ, 4.60^\circ)$.
Fermi J1103.2+1145 is at Galactic coordinates 
 $(l,b) = (238.85^\circ, 60.33^\circ)$, 
 $(\alpha_{J2000},\delta_{J2000})= (165.81^\circ , 11.75^\circ)$.
Possible radio counterparts are found for both sources, which show flat radio
spectra similar to other Fermi LAT detected AGN,
and their identifications are discussed.
\end{abstract}


\section{Introduction}\label{sec:Intro}

The Second Fermi LAT source catalog \citep{Nolan2012} includes as many as 1,873 sources, but
initial attempts to identify counterparts at other wavelengths resulted in 575 sources 
remaining unidentified.
The 2FGL catalog is based on the first 24 months of LAT observation since its launch in 2008, but
the LAT has now accumulated more than 5 years of high-energy gamma-ray data
almost flawlessly, presenting the possibility of finding new sources
which were too faint to be detected in the first two years of data or showed flaring
activity after the catalog was created.

In this paper we report on two new gamma-ray sources serendipitously discovered
in the constellation Leo and discuss possible counterparts based on radio observations.

\section{Analysis}\label{sec:Analysis}

Our original aim was to search for gamma-ray emission from NGC 3628 (Arp 317), one
of the three galaxies called the \lq Leo Triplet', for which
possible starburst activity has been reported based on XMM observations \citep{Tullman2006}.
Five years of archival data of Fermi LAT has been analyzed using the
Fermi Science Tools supplied by Fermi Science Support Center 
(\citet{FSSC}, Fermi Science Tools v9r23p1).
The energy range used in the present analysis was
from 100 MeV to 300 GeV. 
\lq Source' class events detected at zenith angles smaller 
than $100^\circ$ were used for analysis, assuming 
\lq P7SOURCE\_V6' instrument response function 
along with the standard analysis pipeline suggested by FSSC. 
The significance of gamma-ray signal has been estimated 
by maximum likelihood method with a help of the {\tt gtlike} program 
(which we used in the binned mode) included in the tools. 
The data periods for this studies span from 2008 August 4 to 2012 October 27. 

For NGC 3628 (Arp 317),
the test statistic, $TS$, returned by {\tt gtlike} is consistent with zero,
indicating there is no evidence of gamma-ray emission.
Thus we calculated upper limits to gamma-ray flux from NGC 3628
of $1.4\,(1.3)\times 10^{-9}$~cm$^{-2}$s$^{-1}$, at 95\% C.L., above 100\,MeV
for the period 2008 August 4 to 2010 July 31 (2010 July 31 to 2012 October 27).
This is translated to gamma-ray luminosity upper limit of $2.5\,(2.3)\times 10^{43}$~erg~s$^{-1}$
assuming the distance of 12\,Mpc which is derived as the median of 8 measurements,
which range from 6.7 to 14.2\,Mpc \citep{NED}.

\begin{figure}
   \begin{center}
    \includegraphics[width=10cm]{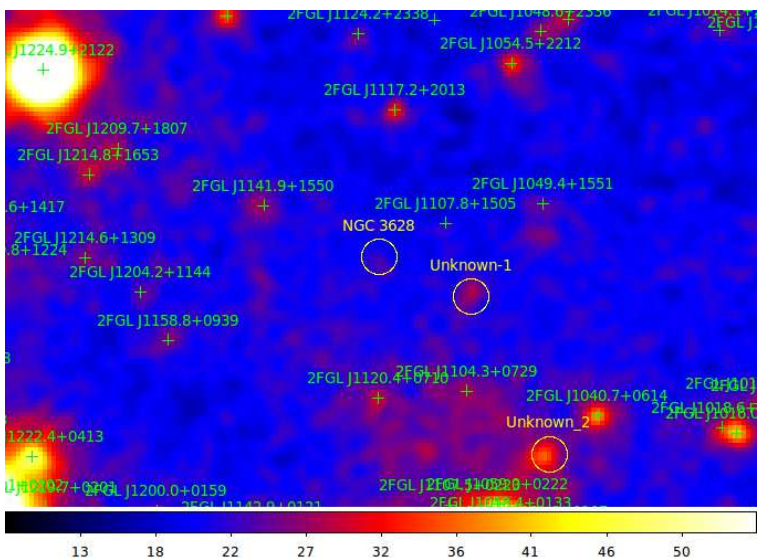}
   \end{center}
 \caption{Gamma-ray countmap around the NGC 3628 region. The map is created
 in $0.1^\circ$ grid and smoothed for the data during 2008 August 05 to 2013 July 03.
  2FGL sources are annotated, and two new gamma-ray
 sources are marked as \lq Unknown\_1' (J1103.2+1145) and \lq Unknown\_2' (J1049.7+0435).}
 \label{fig:cmap}
\end{figure}

\begin{figure}
   \begin{center}
    \includegraphics[width=7cm]{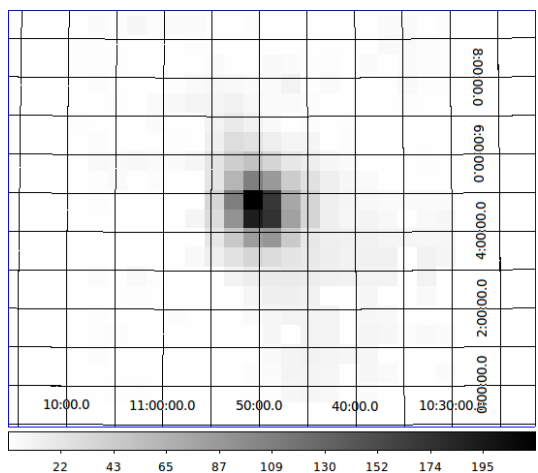}
   \end{center}
 \caption{Test-statistics map around the J1049.7+0435 region for the period
 from 2011 October 27 to 2012 April 27, where the TS value is the highest ($TS=222$).}
 \label{fig:tsmap-2}
\end{figure}

\begin{figure}
   \begin{center}
    \includegraphics[width=7cm]{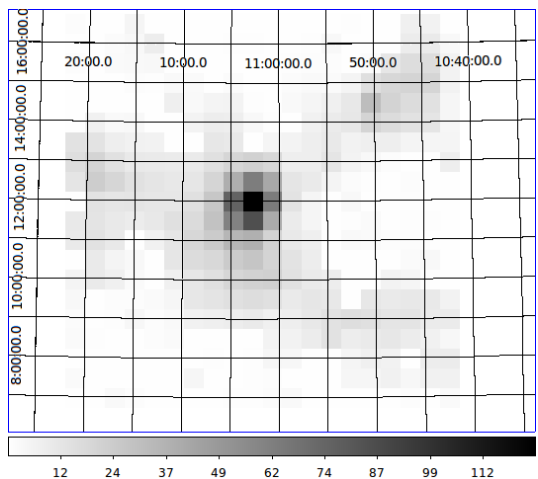}
   \end{center}
 \caption{Test-statistics map around the J1103.2+1145 region for the period
 from 2013 April 27 to 2013 October 27, where the TS value is the highest ($TS=197$).}
 \label{fig:tsmap-1}
\end{figure}

During the study of NGC 3628, we noticed two rather bright gamma-ray sources in the
field of view centered on NGC 3628 and within a radius of $15^\circ$ \citep{ATel}.
They are not coincident with any source listed in the 2FGL catalogue \citep{Nolan2012}
nor in the 3EG catalogue \citep{3EG}.
Figure \ref{fig:cmap} shows a gamma-ray countmap of this area.
The positions for these sources were estimated using the {\tt gttsmap} program
which calculates the $TS$ value assuming an unknown source at various positions
in the field-of-view of interest, and the maximum $TS$ values were obtained 
for positions shown in the Table \ref{tab:pos}. The errors of the positions
are conservatively estimated as the radius at which the $TS$ value drops 
to the half value.
Figures \ref{fig:tsmap-2} and \ref{fig:tsmap-1} show the TS maps around the new
sources for the half-year period where the TS value takes the maximum, which
are consistent with the assumption of point sources.
The average fluxes, the power-law indices, and the TS value for the whole analyzed period
(fron 2008 August 27 to October 10, 2013) are:
$(3.68\pm 0.36)\times 10^{-8}$~cm$^{-2}$s$^{-1}$ ($>100$~MeV), $-2.60\pm 0.07$ and
$TS=332$ for J1049.7+0435;
$(2.35\pm 0.30)\times 10^{-8}$~cm$^{-2}$s$^{-1}$ ($>100$~MeV), $-2.55\pm 0.09$ and
$TS=196$ for J1103.2+1145.

\begin{table}[htbp]
\caption{Best positions of new sources}
\label{tab:pos}
\begin{center}
\begin{tabular}{|cccccc|} \hline
Name & $\alpha_{J2000}$ (deg) & $\delta_{J2000}$ (deg) & $\ell^{\rm II}$ (deg) & $b^{\rm II}$ (deg) & error radius (arcmin) \\ \hline 
J1049.7+0435 & 162.43 &  4.60 & 245.34 & 53.27 & 51 \\ 
J1103.2+1145 & 165.81 & 11.75 & 238.85 & 60.33 & 66 \\ \hline 
\end{tabular}
\end{center}
\end{table}

\begin{figure}
   \begin{center}
    \includegraphics[width=10cm]{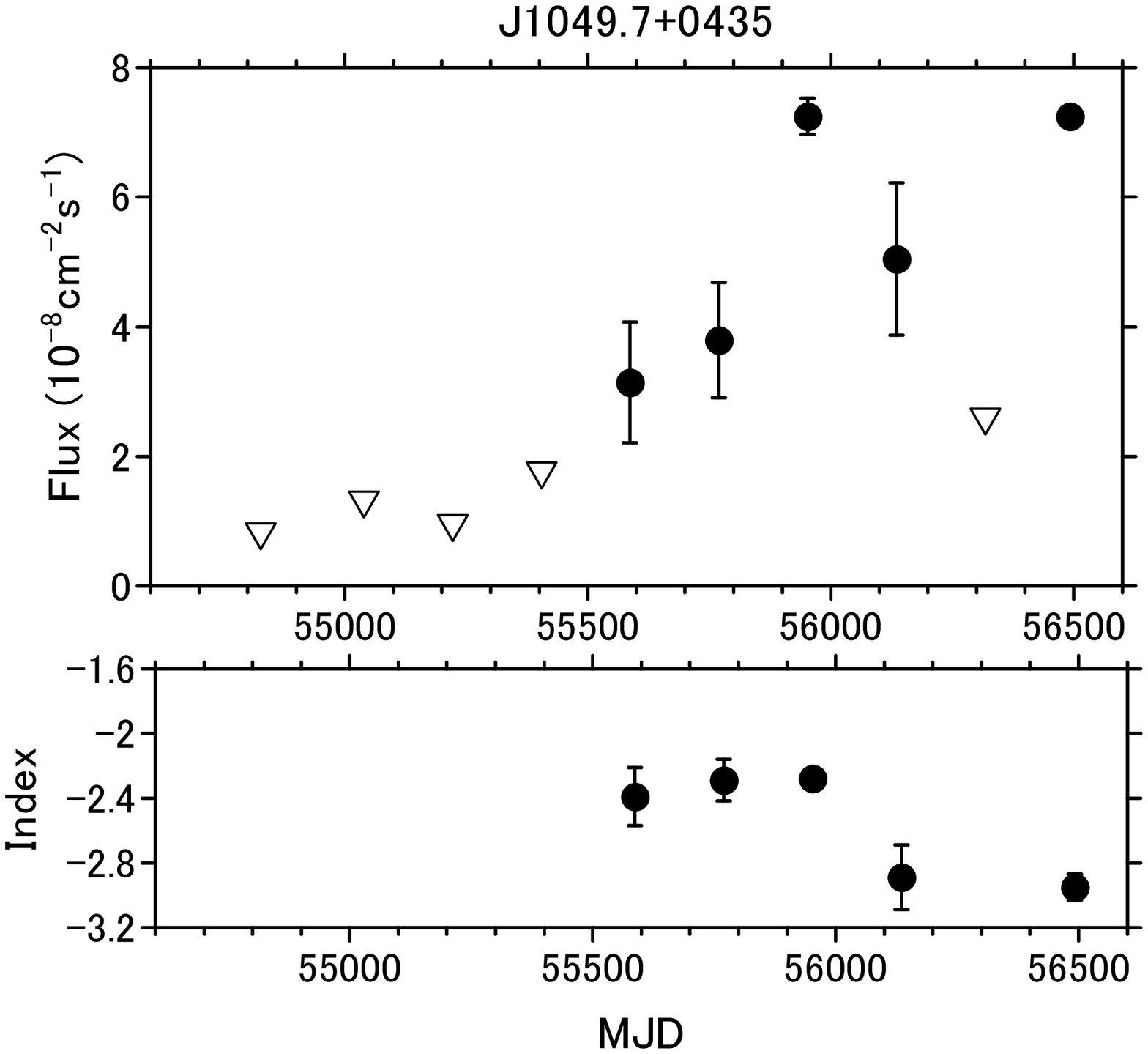}
   \end{center}
 \caption{Time variation of gamma-ray flux of J1049.7+0435 in half-year bins
for the period from 2008 August 5 to 2013 October 10.
 Triangles are upper limits (95\% C.L.).}
 \label{fig:timevar1}
\end{figure}

\begin{figure}
   \begin{center}
    \includegraphics[width=10cm]{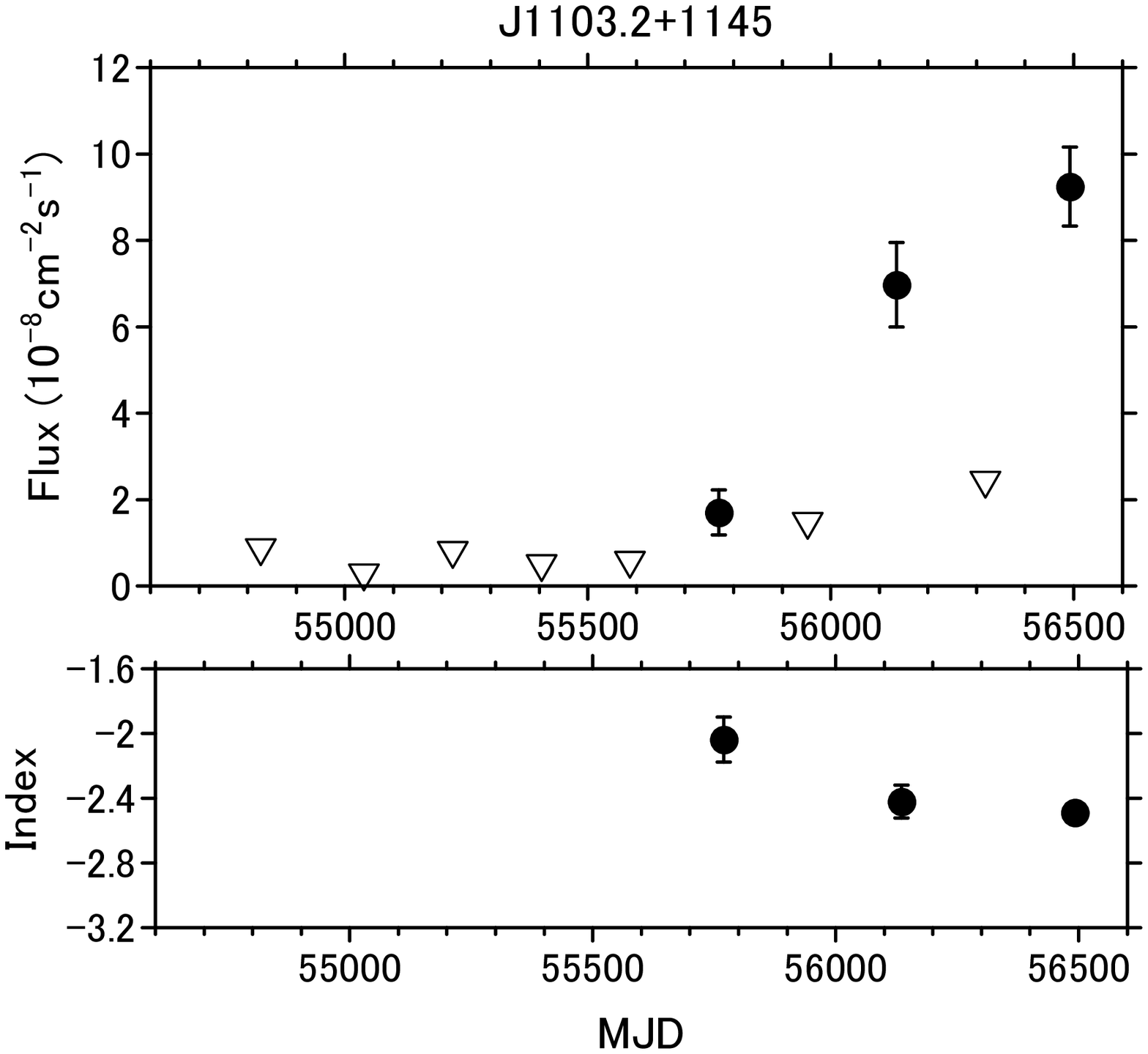}
   \end{center}
 \caption{Time variation of gamma-ray flux of J1103.2+1145 in half-year bins
for the period from 2008 August 5 to 2013 October 10.
 Triangles are upper limits (95\% C.L.).}
 \label{fig:timevar2}
\end{figure}

Figures \ref{fig:timevar1} and \ref{fig:timevar2} show the time variation
of gamma-ray fluxes of the newly detected sources in half-year bins.
For these plots we added data until 2013 October 10.
One can see in the first two years their fluxes are below the
detection threshold ($TS<25$), which is why they are not listed in the 2FGL catalog
based on data over the similar period \citep{Nolan2012}.
The first of our two sources has subsequently been
detected in the Fermi All-sky Variability Analysis \citep{FAVA}
and catalogued as 1FAV J1051+04.

\section{Discussion}\label{sec:Discussion}

The variability of both sources display at gamma-ray energies 
suggests they are more likely to be AGN than members of
other populations of identified gamma-ray sources
\citep{nol03}. The gamma-ray spectral indices are
more consistent with those of flat spectrum radio quasars (FSRQs)
than of BL Lac objects \citep{2LAC}: FSRQs are on average found to be 
more variable than BL Lac objects \citep{nol03,2LAC}.

\citet{mat97,mat01} showed that (extragalactic) 
gamma-ray sources were more likely to be associated
with brighter radio sources, and in particular those sufficiently
compact to be detectable in VLBI observations. 
Radio compactness is generally associated with a flatter
radio spectrum, resulting from synchtrotron self-absorption
at lower frequencies, which is also a characteristic of
Fermi-detected AGN \citep{Abdo2010}.
(Although, as noted by, e.g., \citet{edw05}, radio spectral
indices determined from single dish observations are affected 
by steeper-spectrum radio lobes in some sources, which disguise
the presence of a flat-spectrum radio core.)

We have, therefore, searched for potential counterparts
in the Green Bank 6-cm (GB6, \citet{GB6}) catalog
and determined spectral indices between 20 cm and 6cm
using the NRAO VLA Sky Survey (NVSS, \citet{NVSS}) catalog.

The closest GB6 radio source to J1049.7+0435
is GB6 J1050+0432, with an angular separation of 7 arcmin.
The source has a flux density of 99$\pm$10\,mJy at 4.8 GHz,
and the corresponding 20\,cm source, NVSS J105010+043251,
has a flux density of 101.2\,mJy, yielding a
spectral index $\alpha$ (where $S \propto \nu^{+\alpha}$) 
of 0.0. Two fractionally brighter GB6 sources have both larger 
angular offsets and signficantly {\bf steeper} spectra:
GB6 J1049+0505, 113\,mJy, 30 arcmin separation, $\alpha$=$-$0.8;
GB6 J1051+0449, 101\,mJy, 29 arcmin separation, $\alpha$=$-$0.9.
We note that the GB6 and NVSS flux densities were made some years
apart, and so these spectral indices should be taken as
representative values rather than absolute measurements.
As this declination range is also covered by the 
Parkes-MIT-NRAO equatorial survey \citep{PMNe}, we can compare 
the GB6 value with that
PMN J1050+0432, which has a 4.8\,GHz flux density of 98$\pm$12\,mJy.

For J1105.2+1145, the closest GB6 source is 
GB6 J1103+1158, with an angular separation of 14 arcmin.
The source has a 4.8\,GHz flux density of 306$\pm$27\,mJy,
with the corresponding 20cm source, NVSS J110303+115816,
having a flux density of 262.6\,mJy, resulting in a spectral
index of 0.1. Other relatively bright GB6 sources in the area
are further away and with steeper spectral indices:
GB6 J1103+1114, 116\,mJy, 31 arcmin, $\alpha$=$-$0.7;  
GB6 J1104+1103, 277\,mJy, 46 arcmin, $\alpha$=$-$0.8.  
A Seyfert 1 galaxy, Mrk 728, is 0.89 deg from J1103.2+1145 and 
is not likely the counterpart. 
GB6 J1103+1158 corresponds to the quasar
SDSS 110303.52+115816.5, which lies at a redshift of
0.912 \citep{sch07}.
Furthermore, the quasar has been detected in the VLBA
Calibrator Survey VLBI observations \citep{VC3}, confirming
the presence of a compact core in this radio-loud AGN.

Catalogued radio positions and flux densities 
for the two sources are tabulated in Table~\ref{tab:radio}.

\begin{table}[htbp]
\caption{Possible radio counterparts,}
\label{tab:radio}
\begin{center}
\begin{tabular}{|c|cccc|} \hline
Gamma-ray & catalog     & RA (J2000) & Dec (J2000) & Radio flux    \\ 
source    & (frequency) &            &             & density (mJy) \\ \hline 
J1049.7+0435
  & NVSS (1.4 GHz) & 10 50 10.06 & +04 32 51.3 & 101.2 \\
  & GB6  (4.8 GHz) & 10 50 08.6  & +04 32 37   & 99 \\ \hline
J1103.2+1145
  & NVSS (1.4 GHz) & 11 03 03.55 & +11 58 16.6 & 262 \\
  & GB6  (4.8 GHz) & 11 03 03.7  & +11 58 20   & 306 \\ \hline 
\end{tabular}
\end{center}
\end{table}

We have additionally made snap-shot observations of J1049.7+0435 and
J1103.2+1145 (at their NVSS positions) with the Australia Telescope
Compact Array at several epochs, as part of an on-going program to
monitor gamma-ray sources \citep{C1730} with the measured flux
densities are listed in Table \ref{tab:atca}.  
The observations at 17 GHz and 38 GHz were preceded by a
pointing scan on a nearby bright compact source to refine the global
pointing model.  Data were processed in Miriad in the standard manner.
Flux density calibration was bootstrapped to the standard ATCA flux 
density calibrator, PKS 1934$-$638.
Errors are conservatively estimated as 5\% at lower frequencies and
10\% at highest frequencies, where these include statistical and
systematic errors, with the latter dominating.  

GB6 J1050+0432 has brightened considerably, by a factor of 2.7, 
since the GB6 and PMN observations (which date back to the late 1980s
and early 1990s), and has an inverted spectrum with $\alpha \sim 0.25$,
strengthening the case for an association with J1049.7+0435.
Note also the increased gamma-ray flux in the latest half year 
(Fig.\ref{fig:timevar1}).

GB6 J1103+1158 is a little fainter than the catalogued GB6 value,
however the ATCA observations
confirm that the spectral index remains flat, at $\alpha \sim -0.1$, up
to 38\,GHz. There is no evidence of significant variability over the 4
months spanned by these observations, however comparison with the GB6
flux density indicates the presence of longer timescale variability.

\begin{table}[htbp]
\caption{ATCA radio observations (unit: mJy). See text for details.}
\label{tab:atca}
\begin{center}
\begin{tabular}{|c|cccccc|} \hline
Gamma-ray source & Epoch & Array & 5.5 GHz & 9.0 GHz & 17 GHz  & 38 GHz  \\ \hline
GB6 J1050+0432 & 2013 Oct 20 & H214 & 276 & 311 & 371 &      \\ \hline
GB6 J1103+1158 & 2013 May 10 & 6C   & 254 & 238 & 216 & 230  \\ \cline{2-7} 
               & 2013 Aug 20 & H168 & 254 & 245 &    &       \\ \cline{2-7} 
               & 2013 Sep 8  & 1.5A & 246 & 230 &    &       \\ \hline 
\end{tabular}
\end{center}
\end{table}

We have also examined the ASDC Sky Explorer (ASDC, \citet{ASDC}) and
NASA/IPAC Extragalactic Database (NED, \citet{NED}) for other possible
counterparts, but we did not find any good candidates nearer than
radio sources mentioned above.

In the light of the above facts, we tentatively identify both gamma-ray
sources with the radio sources mentioned above.  
\citet{AOFUS} make a detailed consideration of the utility of
radio observations in finding counterparts to unidentified Fermi sources.
The associations proposed here
would be strengthened by improved gamma-ray localisations, and/or
evidence of contemporaneous multi-wavelength flaring, and, in the case of
GB6 J1050+0432, with VLBI observations to determine whether the source
contains a compact, parsec-scale, radio core.

\section{Conclusions}\label{sec:Conclusions}

A search for gamma-rays from NGC 3628 (Arp 317), for which possible
starburst activity has been reported, found no evidence for
$>$100\,MeV emission.  However, two new GeV sources, Fermi
J1049.7+0435 and J1103.2+1145, have been found near the Leo Triplet
region using Fermi-LAT archival data spanning 5 years.  The fluxes for
both sources increase over the 5 yr period: thus they are not included
in 2FGL catalog.  Their flux variability and spectral indices are
compatible with those of gamma-ray detected AGN.  Based on angular
separation, radio flux density and spectral index, we associate
J1049.7+0435 with GB6 J1050+0432, and J1103.2+1145 with the quasar GB6
J1103+1158.  Further multiwavelength studies are required to confirm
these identifications.

Finally, looking forward to the release of the third Fermi LAT catalog,
we note that the 15 deg radius (0.2 sr) field studied here 
has yielded two previously uncataloged sources.
Although the small number statistics result in extrapolations 
having large uncertainties, this suggests 
(ignoring further improvements in analysis software and background models) 
there might be $\sim$100 new extra-galactic sources
with $|b|>10^\circ$ in the 3FGL catalog.

\acknowledgments

This work is supported in part by the Grant-in-Aid 
from the Ministry of Education, Culture, Sports, Science
and Technology (MEXT) of Japan, No. 22540315.
The Australia Telescope Compact Array is part of the Australia
Telescope National Facility which is funded by the Commonwealth of
Australia for operation as a National Facility managed by CSIRO.
Leonid Petrov is acknowledged for his independent suggestion of
GB6 J1103+1158 as the possible counterpart to
J1103.2+1145.

\end{document}